# Wide-angle giant photonic spin Hall effect

Zhihao Chen[1,2,†], Yu Chen[2,†], Yaodong Wu[1], Xinxing Zhou[1,*], Handong Sun[3,4,*], Tony Low[5], Hongsheng Chen[6,7], and Xiao Lin[6,7,*]

[1]*Key Laboratory of Low-Dimensional Quantum Structures and Quantum Control of Ministry of Education, Synergetic Innovation Center for Quantum Effects and Applications, School of Physics and Electronics, Hunan Normal University, Changsha 410081, China*

[2]*International Collaborative Laboratory of 2D Materials for Optoelectronics Science and Technology, Engineering Technology Research Center for 2D Material Information Function Devices and Systems of Guangdong Province, Institute of Microscale Optoelectronics, Shenzhen University, Shenzhen 518060, China*

[3]*Division of Physics and Applied Physics, School of Physical and Mathematical Sciences, Nanyang Technological University, Singapore 637371, Singapore*

[4]*Centre for Disruptive Photonic Technologies (CDPT), School of Physical and Mathematical Sciences, Nanyang Technological University, Singapore 637371, Singapore*

[5]*Department of Electrical and Computer Engineering, University of Minnesota, Minneapolis, Minnesota 55455, USA.*

[6]*Interdisciplinary Center for Quantum Information, State Key Laboratory of Modern Optical Instrumentation, ZJU-Hangzhou Global Scientific and Technological Innovation Center, College of Information Science and Electronic Engineering, Zhejiang University, Hangzhou 310027, China*

[7]*International Joint Innovation Center, ZJU-UIUC Institute, The Electromagnetics Academy at Zhejiang University, Zhejiang University, Haining 314400, China*

[†]*These authors contributed equally to this work*

[*]*Corresponding authors: xinxingzhou@hunnu.edu.cn (X. Zhou); hdsun@ntu.edu.sg (H. Sun); xiaolinzju@zju.edu.cn (X. Lin)*

**Photonic spin Hall effect is a manifestation of spin-orbit interaction of light and can be measured by a transverse shift $\delta$ of photons with opposite spins. The precise measurement of transverse shifts can enable many spin-related applications, such as precise metrology and optical sensing. However, this transverse shift is generally small (i.e. $\delta/\lambda < 10^{-1}$, $\lambda$ is the wavelength), which impedes its precise measurement. To-date proposals to generate giant spin Hall effect (namely with $\delta/\lambda > 10^2$) have severe limitations, particularly its occurrence only over a narrow angular cone (with a width of $\Delta\theta < 1^o$). Here we propose a universal scheme to realize the wide-angle giant photonic spin Hall effect with $\Delta\theta > 70^o$ by exploiting the interface between free space and uniaxial epsilon-near-zero media. The underlying mechanism is ascribed to the almost-perfect polarization splitting between $s$ and $p$ polarized waves at the designed interface. Remarkably, this almost-perfect polarization splitting does not resort to the interference effect and is insensitive to the incident angle, which then gives rise to the wide-angle giant photonic spin Hall effect.**

Due to its transverse nature, light can possess the longitudinal spin, which refers to the left-handed (denoted as +) or right-handed (−) circular polarization. In analogy to electronic spin Hall effect, photonic spin Hall effect describes the spin-dependent transverse splitting of a light beam upon reflection at an optical interface [1-3]. This spin-dependent transverse shift of reflected light beams $\delta_\pm$, also known as the Imbert-Fedorov shift, offers a unique probe for the spin-orbit interaction of light [4-16].

Generally, the spin-orbit interaction of light is weak at regular optical interfaces, which oftentimes leads to a small transverse shift for reflected light beams [17]. Typically, we have $|\delta_\pm/\lambda| < 10^{-1}$, where $\lambda$ is the wavelength in free space. Such a small transverse shift would impede the direct measurement of photonic spin Hall effect with high accuracy. As a result, the first experimental observation of photonic spin Hall effect in 2008 had to resort to an additional yet complicated setup of weak measurement [4], which later became the key route for the indirect measurement of photonic spin Hall effect. By contrast, the occurrence of giant transverse shifts can allow us to directly measure the photonic spin Hall effect without the help of weak measurement. Such an advantage then continuously motivates the search for giant photonic spin Hall effects, such as those with $|\delta_\pm/\lambda| > 10^2$. The reported schemes to achieve the giant photonic spin Hall effect mainly exploit the Brewster effect [18,19], hyperbolic metamaterials [20,21], and surface plasmon resonances [22,23]. However, due to the sensitivity of these schemes to the incident angle of light beams, the reported giant photonic spin Hall effect can occur only within a narrow angular range with a width of $\Delta\theta < 1^o$. Currently, the realization of giant photonic spin Hall effect over a wide angular range (e.g., $\Delta\theta > 10^o$) remains an open challenge.

Here we propose a viable route to realize the giant photonic spin Hall effect with $|\delta_\pm/\lambda| > 10^2$ over a wide angular range with $\Delta\theta > 70^o$. The underlying physical mechanism for such a wide-angle giant photonic spin Hall effect is due to the exotic reflection phenomenon of interferenceless polarization splitting at the interface between free space and the uniaxial epsilon-near-zero medium. We identify a particular setup where the spin-dependent transverse shift is mainly determined by the reflection ratio $|r_p|/|r_s|$ between $p$ and $s$ polarized waves. Remarkably, the uniaxial epsilon-near-zero medium can be exploited to achieve the almost-perfect polarization splitting and thus a large reflection ratio (e.g. $|r_p|/|r_s| > 10^2$) for almost arbitrary incident angle, without resorting to any interference effect. Although the phenomenon of interferenceless polarization splitting has been revealed by exploiting nanoscale van der Waals heterostructures in Ref. 24 [24], Ref. 24 has nothing to do with the spin-orbit interaction of light. Our work actually bridges the gap between the interesting reflection/transmission phenomenon of interferenceless polarization splitting and the fancy spin-orbit interaction phenomenon of wide-angle giant photonic spin Hall effect.

We begin with the introduction of the photonic spin Hall effect at the interface between free space and a uniaxial epsilon-near-zero medium, as schematically shown in Fig. 1(a). The uniaxial epsilon-near-zero medium has a relative permittivity of $\bar{\bar{\varepsilon}}_r = [\varepsilon_{||}, \varepsilon_{||}, \varepsilon_\perp]$, along with $\varepsilon_{||} \to 1$ and $\varepsilon_\perp \to 0$. Here we consider the incidence of an *s*-polarized Gaussian beam in Fig. 1(a), whose beam width and incident angle are $w$ and $\theta_i$, respectively. Through plane wave expansion and by enforcing the electromagnetic boundary conditions, the transverse shift for the reflected beams with left-handed circular polarization (namely $\delta_+$) or right-handed circular polarization ($\delta_-$) can be expressed as [25]

$$\delta_\pm = \mp \frac{kw^2|r_s|^2\left[1+\frac{|r_p|}{|r_s|}\cos(\varphi_p-\varphi_s)\right]\cot\theta_i}{k^2w^2|r_s|^2+\left|\frac{\partial r_s}{\partial \theta_i}\right|^2+\left[|r_s|^2+|r_p|^2+2|r_s||r_p|\cos(\varphi_p-\varphi_s)\right]\cot^2\theta_i}, \quad (1)$$

where $r_s = |r_s|e^{i\varphi_s}$ and $r_p = |r_p|e^{i\varphi_p}$ are the reflection coefficients at the designed interface for transverse-electric (TE, or *s*-polarized) or transverse-magnetic (TM, or *p*-polarized) waves, respectively; $k = 2\pi/\lambda$ and $\lambda$ are the wavevector and the wavelength of light in free space, respectively. Upon close inspection of equation (1), when $w/\lambda$ is large enough, the denominator in equation (1) will be dominated by the term of $k^2w^2|r_s|^2$. Under this scenario, the transverse shift $\delta_\pm/\lambda$ would be proportional to $|r_p|/|r_s|$ for arbitrary incident angle. This way, the giant transverse shift is in principle achievable by engineering the large value of $|r_p|/|r_s|$.

As a typical example, the large value of $|r_p|/|r_s|$ can be realized by letting $|r_p| \to 1$ and $|r_s| \to 0$ [Fig. 1(b)]. These critical conditions on reflection coefficients actually correspond to the almost-perfect polarization splitting at the interface; that is, the reflected (transmitted) light beam is dominated by the $p$-polarized ($s$-polarized) waves, irrespective of the polarization of incident light beams. Such an exotic phenomenon of polarization splitting can be realized by using a uniaxial medium with $\varepsilon_\| \to 1$ and $\varepsilon_\perp \to 0$ [25]. Remarkably, Fig. 1(c) shows that we can have $|r_p|/|r_s| > 10^2$ within the angular range of $\theta_i \in [0^o, 89^o]$, for example, by letting $\varepsilon_\| = 1 - 10^{-4}$ and $\varepsilon_\perp = 10^{-4}$. The result in Fig. 1(c) validates that a high value of $|r_p|/|r_s|$ can be achieved in a very wide angular range by exploiting a uniaxial epsilon-near-zero medium.

With this uniaxial epsilon-near-zero medium, the transverse shift is plotted as a function of the incident angle in Fig. 1(d), along with $w/\lambda = 10^4$. Remarkably, the maximum value of $|\delta_\pm/\lambda|$ can approach the order of $10^4$, and we have $|\delta_\pm/\lambda| > 10^2$ within the broad angular range of $\theta_i \in [\theta_{min}, \theta_{max}]$, with $\theta_{min} = 1.1^o$ and $\theta_{max} = 75.6^o$. In other words, the angular width for the occurrence of the giant photonic spin Hall effect is $\Delta\theta = \theta_{max} - \theta_{min} = 74.5^o$. Therefore, the exotic phenomenon of almost-perfect polarization splitting provides a feasible mechanism for the realization of the wide-angle giant photonic spin Hall effect. For comparison, we also investigate the reflection and the transverse shift at a regular interface between free space and glass in Fig. 1(c-d). By contrast, at this regular interface, we always have $|r_p|/|r_s| < 1$ in Fig. 1(c) for arbitrary incident angle. This is accompanied with a maximum value of $|\delta_\pm/\lambda|$ of the order $10^{-1}$ in Fig. 1(d).

We emphasize that if the uniaxial medium has $\varepsilon_\| = 1$ (instead of $\varepsilon_\| \to 1$), we would have $|r_s| = 0$ (instead of $|r_s| \to 0$). Under this scenario, the numerator in equation (1) is always zero, which is distinctively different from the physical situation described in Fig. 1. Then according to equation (1), we have $|\delta_\pm/\lambda| = 0$ for arbitrary incident angle, indicating the disappearance of giant photonic spin Hall effect. As such, the uniaxial epsilon-near-zero medium should be judiciously designed in order to obtain the wide-angle giant photonic spin Hall effect.

To facilitate further understanding, Fig. 2 systematically shows the influence of the intrinsic property of the uniaxial epsilon-near-zero medium on the transverse shift. Figure 2(a-b) shows the influence of $\varepsilon_\|$ and $\varepsilon_\perp$ on the wide-angle giant photonic spin Hall effect, respectively. Remarkably, the existence of wide-angle giant photonic spin Hall effect is robust with respect to the variation of either $\varepsilon_\parallel$ or $\varepsilon_\perp$ as shown in Fig. 2, although the dependence of the transverse shift (such as the angular position of its maximum value) is sensitive to these variations.

On the other hand, when we plot $|\delta_\pm/\lambda|$ as a function of the incident angle, a sharp dip emerges for some specific values of $\varepsilon_\perp$, as shown in Fig. 2(b). Here, the angular position of this dip is denoted as $\theta_{\text{dip}}$. To be specific, if $\theta_i = \theta_{\text{dip}}$, we have $|\delta_\pm/\lambda| = 0$. As such, the giant photonic spin Hall effect would also disappear at a very narrow angular range around $\theta_{\text{dip}}$, whose angular width is smaller than $1^o$(shown in APPENDIX A). This is a consequence of the fact that the sign of $\delta_\pm$ at $\theta_i < \theta_{\text{dip}}$ is opposite to that at $\theta_i > \theta_{\text{dip}}$ [Fig. S2(b)], namely the direction of transverse shift would switch at $\theta_i = \theta_{\text{dip}}$. The sharp variation of this dip might be of interest in sensing applications. For example, we show in Fig. S3 that if $\varepsilon_\parallel$ and $\varepsilon_\perp$ are fixed, $\theta_{\text{dip}}$ is extremely sensitive to variation of the refractive index for the region (such as free space filled with different densities of gases) above the uniaxial epsilon-near-zero medium.

From equation (1), it should be noticed that when $w$ is large enough, the contribution from the terms of $\left|\frac{\partial r_s}{\partial \theta_i}\right|^2$ and $\left[|r_s|^2 + |r_p|^2 + 2|r_s||r_p|\cos(\varphi_p - \varphi_s)\right]\cot^2\theta_i$ can be reduced to negligible in the denominator. Therefore, we will discuss the influence of the beam width on the wide-angle giant photonic spin Hall effect. In Fig. 3, the normalized beam width $w/\lambda$ decreases from $\infty$, $10^5$, $10^4$ to $10^3$, while $\varepsilon_\parallel = 1 - 10^{-4}$ and $\varepsilon_\perp = 10^{-4}$ are chosen. Broadly speaking, the transverse shift would increase with $w$, so does the angular width $\Delta\theta$ for the wide-angle giant photonic spin Hall effect. Moreover, we always have $\max(|\delta_\pm|) \leq w/2$, as critically proved in the APPENDIX B.

Remarkably, the wide-angle photonic spin Hall effect is accessible with the usage of a reasonably large beam width, as shown in Fig. 3, which should facilitate its experimental observation. For example, if $w/\lambda = 10^3$ which is deemed as a "worst case" focusing in practical experiments, the giant photonic spin Hall effect with $|\delta_\pm/\lambda| > 10^2$ can still hold for $\theta_i \in [27°, 75.6°]$, namely with $\Delta\theta > 48^o$.

So far, the wide-angle giant photonic spin Hall effect is obtained by exploiting the phenomenon of polarization splitting, without resorting to the interference effect. In general, the designed polarization splitting is not perfect at the single interface between the free space and a uniaxial epsilon-near-zero medium. Hence, one might anticipate that weak interference effect would emerge for the reflected light beams if the infinitely-large uniaxial medium is replaced with a corresponding slab with a finite thickness of $d$, as schematically shown in Fig. 4(a). It would be of fundamental interest to investigate the influence of the weak interference on the wide-angle giant photonic spin Hall effect. Figure 4(b) shows the transverse shift as a function of the incident angle, assuming that the uniaxial epsilon-near-zero slab has a subwavelength thickness $d$, such as $d/\lambda = 0.1$, 0.3 or 0.5. From Fig. 4, the dependence of transverse shift on the incident angle is sensitive to the slab thickness, which originates from various weak interference effects for the reflected light beams. However, the phenomenon of the wide-angle giant photonic spin Hall effect remains robust across this range of slab thickness, which is typical for standard materials growth and preparation.

In our theory, the dielectric tensor of anisotropic media we used is approximately [1, 1, 0]. This anisotropic medium can be practically realized by using graphene-hexagonal boron nitride (h-BN) heterostructure [24]. When the thickness of graphene layer is 0.35nm, h-BN layer is 9nm, the total relative permittivity of graphene-h-BN heterostructure will be approximately [1, 1, 0] at 25.35 THz. As for optical waveband, it can be realized through stacking layers of a plasmonic material and a dielectric material in sub-wavelength scale. By controlling the thickness of plasmonic and dielectric layers, the effective permittivity can be approximately [1, 1, 0].

Last but not least, the uniaxial epsilon-near-zero medium is in principle realizable by following the general design methodology for anisotropic metamaterials [26-29], for example, with the usage of 2D material-based van der Waals heterostructures [25] or metal based multilayer planar structures [30]. In addition, since the material loss is ubiquitous in practical applications, we also evaluate its influence in Fig. S5. Although the performance is degraded, the existence of the wide-angle giant photonic spin Hall effect can still retain under reasonable amount of loss.

In conclusion, we have revealed a viable mechanism to achieve the wide-angle giant photonic spin Hall effect, which originates from the interferenceless phenomenon of polarization splitting at the interface between free space and uniaxial epsilon-near-zero media. The uniaxial epsilon-near-zero medium offers a promising platform to significantly enhance the spin-orbit interaction of light for arbitrary incident angle of light beams. Our finding is thus important for the development of precise metrology, advanced sensing, and spin-based photonic devices. On the other hand, our work further indicates that the exotic photonic spin Hall effect relies heavily on the novel scattering phenomena at the interface. Remarkably, there are other fancy scattering phenomena that are continually being uncovered recently, except for the polarization splitting discussed here. For example, as revealed in Ref. [31], the transmitted light through chiral interfaces, such as atomic bilayers with Moiré superlattice [32,33], can always have polarization different from that of the incident light. Our work then should inspire the search for other exotic photonic spin Hall effects in emerging material systems [30,34-37], such as those with Moiré superlattices [38-40].

**Acknowledgment**
X. Lin was supported by the Fundamental Research Funds for the Central Universities (2021FZZX001-19) and Zhejiang University Global Partnership Fund. X. Zhou was supported by the National Natural Science Foundation of China (11604095) and the Training Program for Excellent Young Innovators of Changsha (kq2107013). H. Sun acknowledges support from Singapore Ministry of Education AcRF Tier 1 (RG95/19 (S)).

**APPENDIX A: MORE DISCUSSION ON THE SHARP DIP IN FIG. 2(b)**

From equation (1), the transverse spin shift is closely related to the reflection coefficients of TE (s-polarized) and TM (p-polarized) waves, which are denoted as $r_s$ and $r_p$, respectively. By enforcing the boundary conditions for electromagnetic waves, the reflection coefficients at the interface between free space and the uniaxial medium with a relative permittivity of $\bar{\bar{\varepsilon}}_r = [\varepsilon_{\parallel}, \varepsilon_{\parallel}, \varepsilon_{\perp}]$ can be calculated as [24]

$$r_p = \frac{\varepsilon_{\parallel} \cos\theta_i - \sqrt{\varepsilon_{\parallel} - \frac{\varepsilon_{\parallel}}{\varepsilon_{\perp}}\sin^2\theta_i}}{\varepsilon_{\parallel} \cos\theta_i + \sqrt{\varepsilon_{\parallel} - \frac{\varepsilon_{\parallel}}{\varepsilon_{\perp}}\sin^2\theta_i}}, \tag{A1}$$

$$r_s = \frac{\cos\theta_i - \sqrt{\varepsilon_{\parallel} - \sin^2\theta_i}}{\cos\theta_i + \sqrt{\varepsilon_{\parallel} - \sin^2\theta_i}}. \tag{A2}$$

Figure. 5 shows the reflection coefficients as a function of the incident angle by using the structural setup in Fig. 2(b). Figure 5(a) shows that when the incident angle is small than $80^o$, $|r_s| \to 0$ and $|r_p| \to 1$. Then the sign of the transverse shift is mainly affected by the phase difference between two reflection coefficients, namely $\varphi_p - \varphi_s = Arg(r_p) - Arg(r_s)$. When the incident angle increases from zero to the large values, $\varphi_p - \varphi_s$ gradually reduces from 0 to $-\pi$, as shown in Fig. 5(b). When the term $\varphi_p - \varphi_s$ decreases to a value close to $-0.5\pi$, we have $1 + \frac{|r_p|}{|r_s|}\cos(\varphi_p - \varphi_s) = 0$ in equation (1), which results in the zero-value transverse shift. When slightly away from this specific value of $\varphi_p - \varphi_s$ by a small variation of the incident angle, $\delta_+/\lambda$ will change rapidly from negative to positive, since $\frac{|r_p|}{|r_s|}$ can be extremely large. The large variation of $|\delta_\pm/\lambda|$ as a function of the incident angle around $\theta_i = \theta_{dip}$ then leads to the formation of a sharp dip at $\theta_i = \theta_{dip}$ in Fig. 2(b).

As another complementary information for Fig. 5(b), we re-plot Fig. 2(b) by showing $\delta_+/\lambda$, instead of $|\delta_\pm/\lambda|$, as a function of the incident angle in Fig. 6. It can be clearly seen that we have $\delta_+/\lambda = 0$ at $\theta_i = \theta_{dip}$, $\delta_+/\lambda < 0$ if $\theta_i < \theta_{dip}$, and $\delta_+/\lambda > 0$ if $\theta_i > \theta_{dip}$. Moreover, the dip with $|\delta_\pm/\lambda| \leq 10^2$ has a width less than $1^o$ for the cases studied in Fig. 2(b).

Based on the ultra-sharpness of this dip in Fig. 2(b), we propose in Fig. 7 a new type of gas sensors based on the accurate measurement of the angular position of the dip, namely $\theta_{dip}$. Figure 7 shows that $\theta_{dip}$ is sensitive to the slight variation of the refractive index for the region (e.g., free space with gases with different densities) above the uniaxial epsilon-near-zero material. As such, the designed gas sensors can have high sensitivity, which is favored for practical detection.

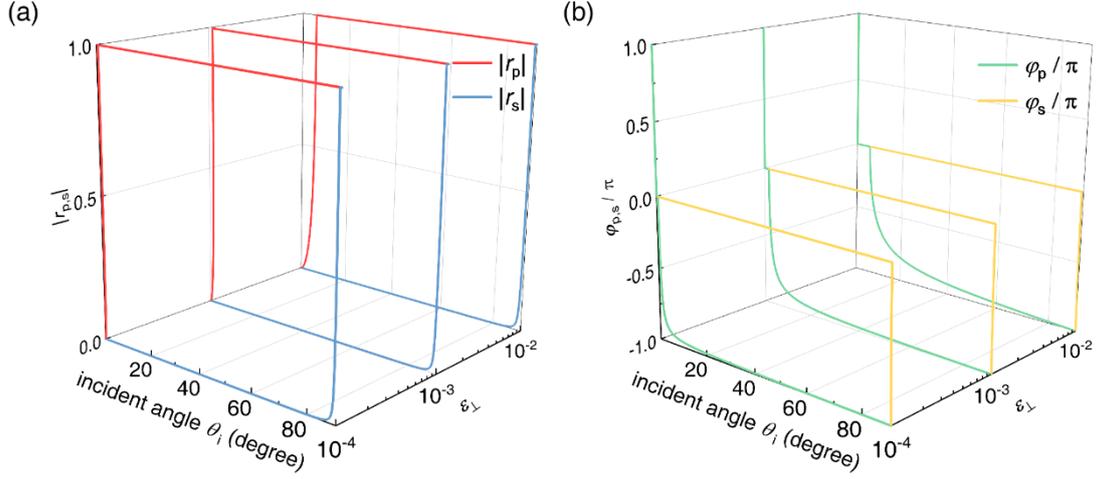

FIG. 5. Reflection coefficients. Here $\varepsilon_\parallel = 1 - 10^{-4}$, and the structural setup is the same as Figure 2b. a) Influence of $\varepsilon_\perp$ on the magnitude $|r_{p,s}|$ of reflection coefficients. b) Influence of $\varepsilon_\perp$ on the phase $\varphi_{p,s} = Arg(r_{p,s})$ of reflection coefficients.

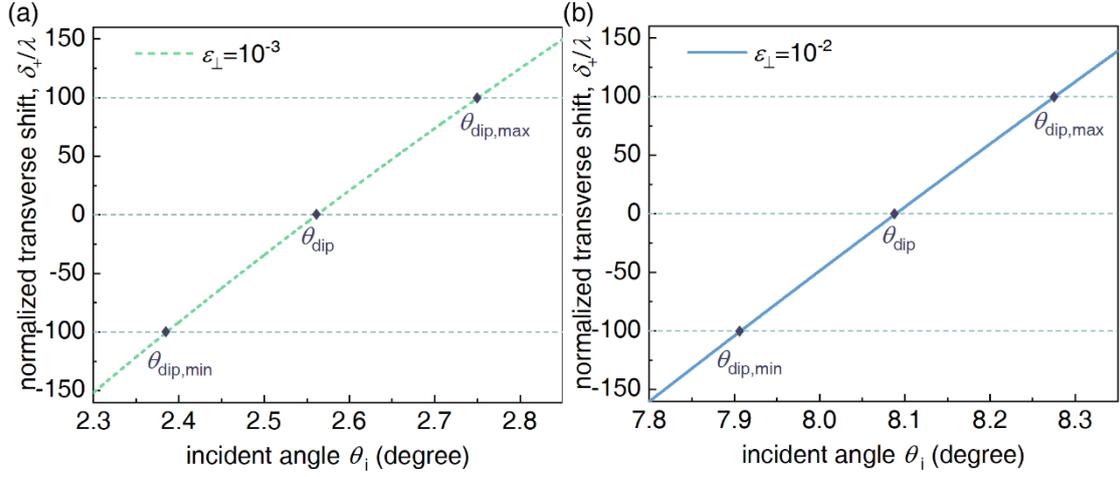

FIG. 6. A close inspection of the dip in Figure 2b. Here we plot $\delta_+/\lambda$, instead of $|\delta_\pm|/\lambda$, as a function of the incident angle. a) $\varepsilon_\perp = 10^{-3}$. b) $\varepsilon_\perp = 10^{-2}$. The other structural setup in (a, b) is the same as Figure 2b. If $\theta_i = \theta_{dip}$, we have $\delta_+/\lambda = 0$. If $\theta_i \in [\theta_{dip,min}, \theta_{dip,max}]$, we have $|\delta_\pm/\lambda| \leq 10^2$. For the cases studied in Figure 2b, we have $\Delta\theta_{dip} = \theta_{dip,max} - \theta_{dip,min} < 1^0$.

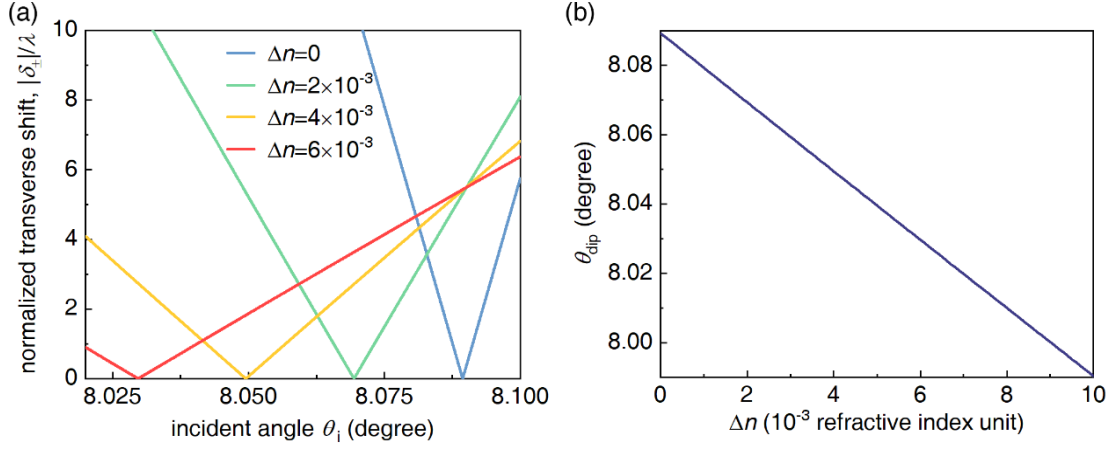

FIG. 7. Gas sensors based on the measurement of the angular position of the dip in Figure 2b. Here $\varepsilon_{\parallel} = 1 - 10^{-4}$ and $\varepsilon_{\perp} = 10^{-2}$. We assume the region (e.g., free space with gases with different densities) above the uniaxial epsilon-near-zero medium has a refractive index of $1 + \Delta n$. The other setup is the same as that is Figure 1d. a) Influence of $\Delta n$ on the transverse shift. b) The angular position of the dip (namely $\theta_{\text{dip}}$) as a function of $\Delta n$. $\theta_{\text{dip}}$ is sensitive to the slight variation of $\Delta n$.

## APPENDIX B: UPPER LIMIT FOR THE TRANSVERSE SHIFT

According to equation (1), the transverse spin shift can be re-organized into the following form

$$|\delta_{\pm}| = \left| \frac{w[|r_s|+|r_p|\cos(\varphi_p-\varphi_s)]\cot\theta_i}{kw|r_s|+\frac{\left|\frac{\partial r_s}{\partial \theta_i}\right|^2}{kw|r_s|}+\frac{[|r_s|^2+|r_p|^2+2|r_s||r_p|\cos(\varphi_p-\varphi_s)]\cot^2\theta_i}{kw|r_s|}} \right|, \tag{B1}$$

Since $\frac{\left|\frac{\partial r_s}{\partial \theta_i}\right|^2}{kw|r_s|} \geq 0$ and $\cos^2(\varphi_p - \varphi_s) \leq 1$, we have the following inequality:

$$\left| \frac{w[|r_s|+|r_p|\cos(\varphi_p-\varphi_s)]\cot\theta_i}{kw|r_s|+\frac{\left|\frac{\partial r_s}{\partial \theta_i}\right|^2}{kw|r_s|}+\frac{[|r_s|^2+|r_p|^2+2|r_s||r_p|\cos(\varphi_p-\varphi_s)]\cot^2\theta_i}{kw|r_s|}} \right| \leq \left| \frac{w[|r_s|+|r_p|\cos(\varphi_p-\varphi_s)]\cot\theta_i}{kw|r_s|+\frac{[|r_s|^2+|r_p|^2\cos^2(\varphi_p-\varphi_s)+2|r_s||r_p|\cos(\varphi_p-\varphi_s)]\cot^2\theta_i}{kw|r_s|}} \right| =$$

$$\frac{w}{\frac{kw|r_s|}{|[|r_s|+|r_p|\cos(\varphi_p-\varphi_s)]\cot\theta_i|}+\frac{|[|r_s|+|r_p|\cos(\varphi_p-\varphi_s)]\cot\theta_i|}{kw|r_s|}} \leq \frac{w}{2}, \tag{B2}$$

Therefore, we always have $|\delta_{\pm}| \leq w/2$.

## APPENDIX C: INFLUENCE OF THICKNESS AND LOSS ON THE TRANSVERSE SHIFT

When considering a slab medium, the reflection coefficient of a slab can be regarded as a three-layers Fresnel model [24]. For conceptual illustration, we assume the uniaxial slab with a thickness of d is surrounded by air. This way, the reflection coefficients need to take multiple scattering into account, and can be regarded as a function of thickness d. The transverse spin shifts can be obtained by substituting these reflection coefficients into equation (1). When the thickness gradually increases, the influence of interference on transverse shift is more intuitive, which is caused by transmitted wave and reflected wave from second interface. From Fig. 8, it can be seen that when the thickness increases from 15λ to 30λ, the zero-reflection point caused by destructively interference increases gradually, which shows the actual dependence of the shift on the epsilon-near-zero slab thickness.

We also show in Fig. 9 that the revealed phenomenon of wide-angle giant photonic spin Hall effect can still exist, if the reasonable amount of material loss is assumed. Here, the slab with loss has a permittivity of $\varepsilon = \varepsilon_R + i\varepsilon_I$. The subscripts represent the real and imaginary parts of the refractive index, where the imaginary part represents the loss of the slab. In combination with the three-layers Fresnel coefficient expression and equation (1), the transverse spin shift can be obtained. In Fig. 9, with the increase of loss, the peak value of spin displacement changes greatly, but the existence of the wide-angle giant photonic spin Hall effect can still retain. Since the imaginary and real parts of the permittivity are of the same order of magnitude, thus we believe that this scheme is still feasible even in a lossy system.

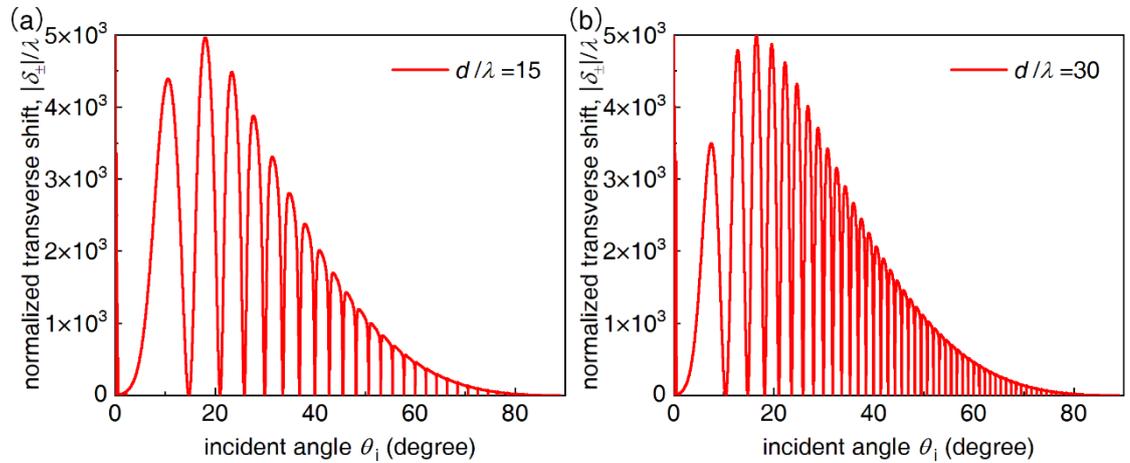

FIG. 8. The influence of slab thickness on the transverse spin shift. For conceptual demonstration, the uniaxial medium has $\varepsilon_{\parallel} = 1 - 10^{-4}$ and $\varepsilon_{\perp} = 10^{-4}$. a) The thickness of slab $d/\lambda = 15$. b) $d/\lambda = 30$.

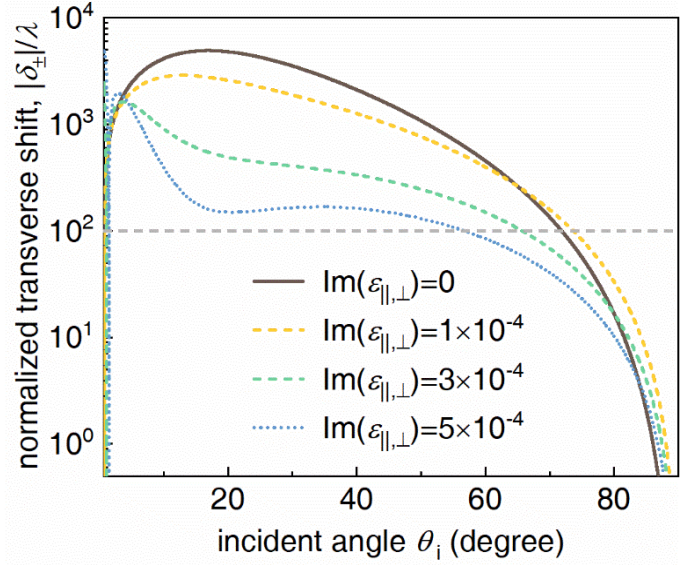

FIG. 9. Loss influence on the transverse shift. This figure severs as the complementary information for Figure 4. The slab thickness of $d/\lambda = 0.3$ is chosen. When considering the loss, we have $\text{Im}(\varepsilon_{\parallel,\perp}) \neq 0$. For conceptual illustration, we set $\text{Im}(\varepsilon_{\parallel}) = \text{Im}(\varepsilon_{\perp})$, $\text{Re}(\varepsilon_{\parallel}) = 1 - 10^{-4}$, and $\text{Re}(\varepsilon_{\perp}) = 10^{-4}$. The other structural setup is the same as Figure 4. When $\text{Im}(\varepsilon_{\parallel,\perp})$ increases, the transverse shift will be decreased. But the reasonable amount of loss would not have a drastic influence on the existence of the wide-angle giant photonic spin Hall effect.

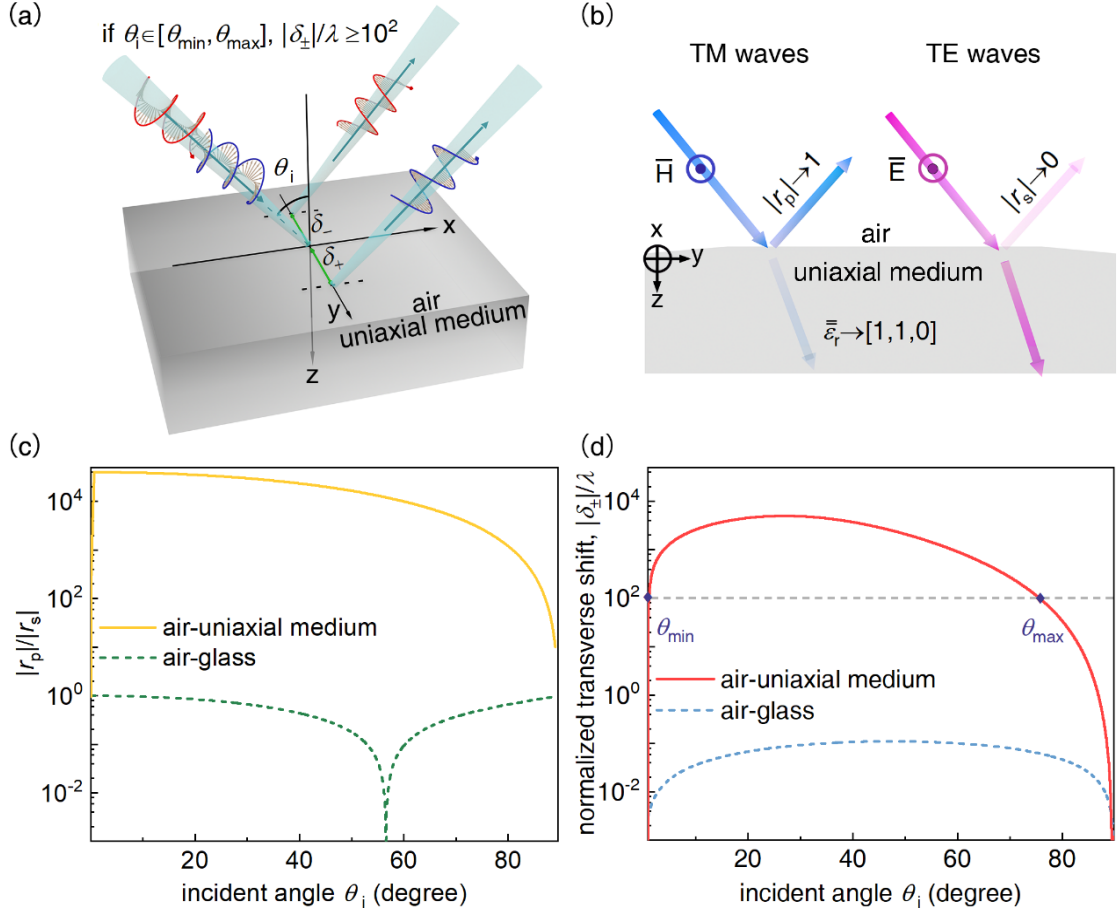

FIG. 1. Wide-angle giant photonic spin Hall effect from a specially designed interface. The interface is between an isotropic medium (e.g. air) and a uniaxial medium with a relative permittivity of $\bar{\bar{\varepsilon}}_r = [\varepsilon_\parallel, \varepsilon_\parallel, \varepsilon_\perp]$, where $\varepsilon_\parallel \to 1$ and $\varepsilon_\perp \to 0$. a) Schematic of the photonic spin Hall effect. This effect is featured with a spin-dependent transverse shift $\delta_\pm$, where the subscript $+$ $(-)$ corresponds to the reflected light with left-handed (right-handed) circular polarization. b) Schematic of reflection. The reflection coefficient for TE (s polarized) or TM (p polarized) waves is denoted as $r_s$ and $r_p$, respectively. c-d) $|r_p|/|r_s|$ and normalized transverse shift $|\delta_\pm|/\lambda$ as a function of the incident angle. In (c-d), $\lambda = 632.8$ nm is the wavelength in free space; $\varepsilon_\parallel = 1 - 10^{-4}$ and $\varepsilon_\perp = 10^{-4}$; the beam width of $w = 10^4 \lambda$ is chosen for the incident s-polarized Gaussian beam. The giant photonic spin Hall effect with $|\delta_\pm|/\lambda \geq 10^2$ occurs within a wide incident angular range, namely from $\theta_{min} = 1.1°$ to $\theta_{max} = 75.6°$ as shown in (d). For comparison, the reflection and the photonic spin Hall effect at the regular air-glass interface are also studied in (c, d), where the refractive index of glass is $n = 1.515$.

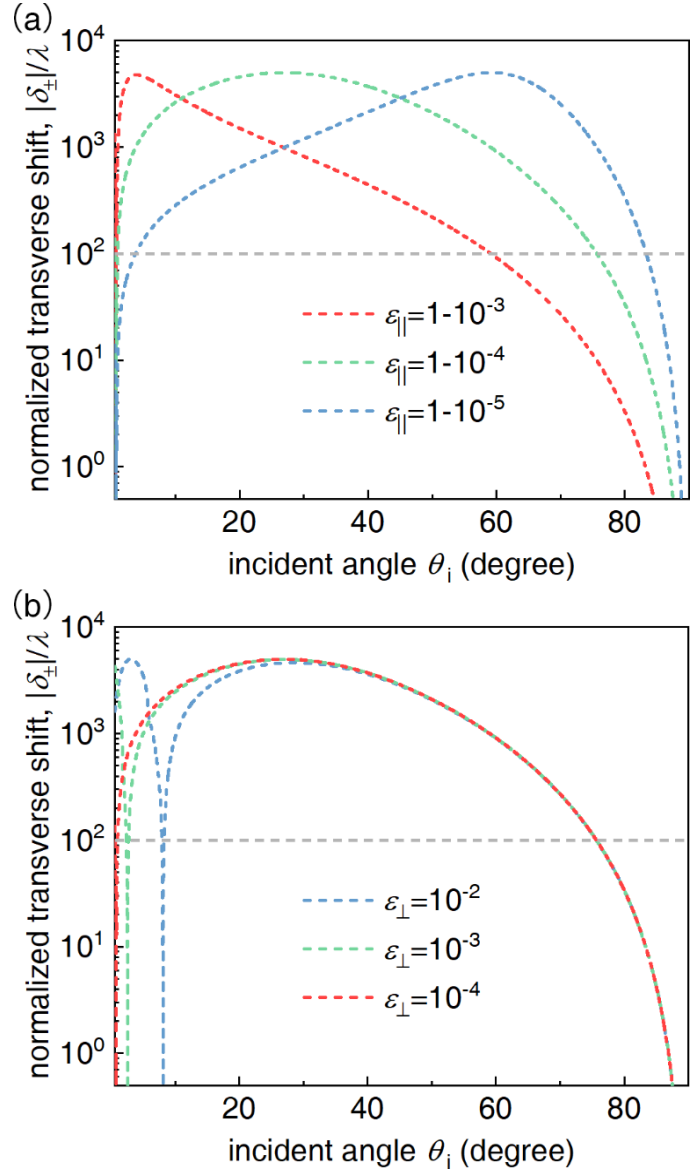

FIG. 2. Influence of the uniaxial medium on the wide-angle giant photonic spin Hall effect at the air-uniaxial medium interface. The uniaxial medium has a relative permittivity of $\bar{\bar{\varepsilon}}_r = [\varepsilon_{||}, \varepsilon_{||}, \varepsilon_\perp]$, where $\varepsilon_{||} \to 1$ and $\varepsilon_\perp \to 0$. a) Influence of $\varepsilon_{||}$ on the transverse shift, under the scenario of $\varepsilon_\perp = 10^{-4}$. b) Influence of $\varepsilon_\perp$ on the transverse shift, by setting $\varepsilon_{||} = 1 - 10^{-4}$.

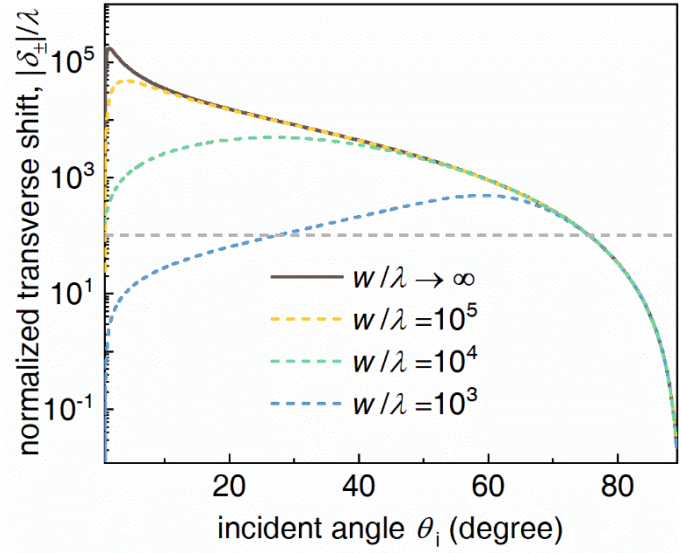

FIG. 3. Influence of the beam width on the wide-angle giant photonic spin Hall effect. The other structural setup is the same as Fig. 1(d), except for the width $w$ of the incident Gaussian beam.

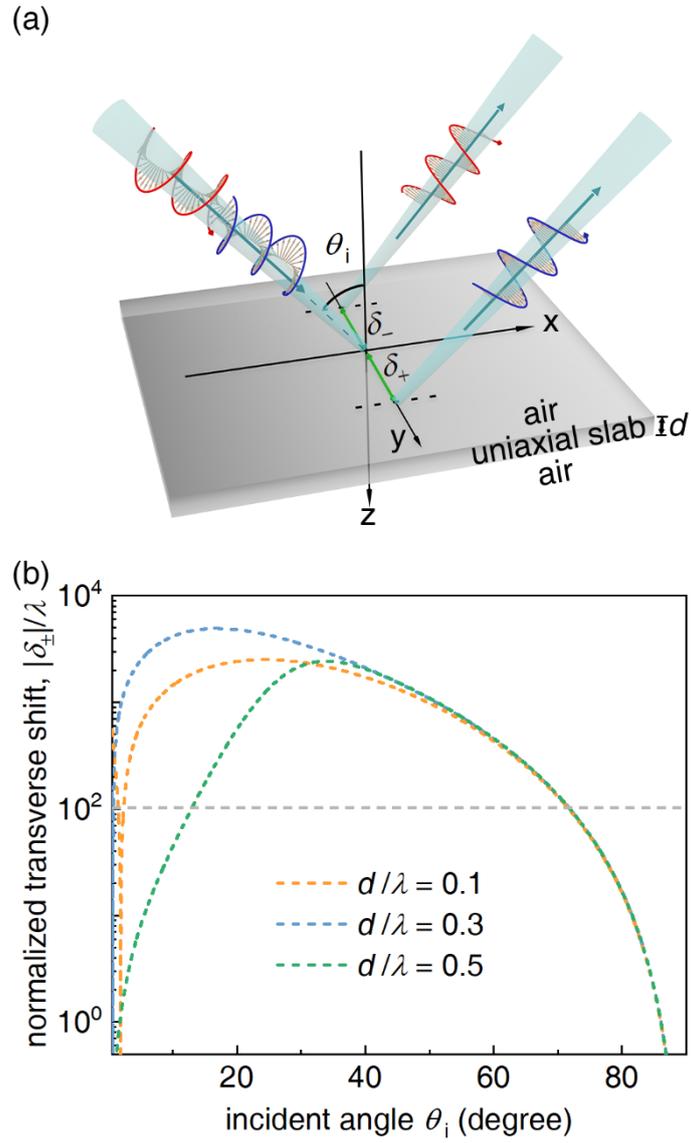

FIG. 4. Wide-angle giant photonic spin Hall effect from uniaxial slabs with a subwavelength thickness. a) Schematic of the photonic spin Hall effect. The surrounding environment is free space. The other structural setup is the same as Fig. 1(d), except for the slab thickness $d$. b) Influence of the slab thickness on the transverse shift.